\documentclass[conference]{IEEEtran}
\IEEEoverridecommandlockouts
\usepackage{cite}
\usepackage{amsmath,amssymb,amsfonts}
\usepackage{booktabs}
\usepackage{algorithmic}
\usepackage{graphicx}
\usepackage{textcomp}
\usepackage{xcolor}
\usepackage{latexsym}
\usepackage [english]{babel}
\usepackage [autostyle, english = american]{csquotes}
\usepackage{placeins}

\MakeOuterQuote{"}
\def\BibTeX{{\rm B\kern-.05em{\sc i\kern-.025em b}\kern-.08em
    T\kern-.1667em\lower.7ex\hbox{E}\kern-.125emX}}
\begin{document}

\title{  Anisotropic Diffusion Model of Communication in 2D Biofilm}

\author{\IEEEauthorblockN{Yanahan Paramalingam, Hamidreza Arjmandi, Freya Harrison, \\
Tara Schiller, and Adam Noel, \IEEEmembership{Senior Member,~IEEE}}

\thanks{This work was supported by the Engineering and Physical Sciences Research Council grant number EP/V030493/1. For the purpose of open access, the authors have applied a Creative Commons Attribution (CC-BY) licence to any Author Accepted Manuscript version arising from this submission.} 

\thanks{Y.~Paramalingam and A.~Noel are with the School of Engineering, University of Warwick, Coventry, UK. (e-mail: \{yanahan.paramalingam.1, Adam.Noel\}@warwick.ac.uk)}

\thanks{H.~Arjmandi is with the Institute of Cancer and Genomic Sciences, University of Birmingham, Birmingham, UK. (e-mail: h.arjmandi@bham.ac.uk.)}

\thanks{F.~Harrison is with the School of Life Sciences, University of Warwick, Coventry, UK. (e-mail: f.harrison@warwick.ac.uk)}

\thanks{T.~Schiller is with the Warwick Manufacturing Group, University of Warwick, Coventry, UK. (e-mail: t.l.schiller@warwick.ac.uk)}
}

\maketitle

\begin{abstract}
A biofilm is a microbial city. It consists of bacteria embedded in an extracellular polymeric substance (EPS) that functions as a protective barrier. Quorum sensing (QS) is a method of bacterial communication, where autoinducers (AIs) propagate via diffusion through the EPS and water channels within the biofilm. This diffusion process is anisotropic due to varying densities between the EPS and water channels. This study introduces a 2D anisotropic diffusion model for molecular communication (MC) within biofilms, analyzing information propagation between a point-to-point transmitter (TX) and receiver (RX) in  bounded space. The channel impulse response is derived using Green's function for concentration (GFC) and is validated with particle-based simulation (PBS).
The outcomes reveal similar results for both isotropic and anisotropic diffusion when the TX is centrally located due to symmetry. However, anisotropic conditions lead to greater diffusion peaks when the TX is positioned off-center. Additionally, the propagation of AIs is inversely proportional to both overall biofilm size and and diffusion coefficient values. It is hypothesized that anisotropic diffusion supports faster responses to hostile environmental changes because signals can propagate faster from the edge of the biofilm to the center.
\end{abstract}


\section{Introduction}
 
Bacteria can adapt their behaviors to environmental conditions. Colonies of bacteria have the capacity of multicellular organization and cellular differentiation, which can lead to the formation of microbial cities called biofilms \cite{hall2004bacterial,costerton1999bacterial}. Biofilms are aggregations of bacteria buried in or attached to the surface of an extracellular polymeric substance (EPS). Within a biofilm, substances such as extracellular DNA, amyloidogenic proteins, polysaccharides, and proteins can be found that enable the biofilm to withstand harsh conditions \cite{vestby2020bacterial}. In healthcare, biofilm formation is a major concern due to them having multidrug resistance and having the ability to withstand other external stresses, thereby contributing to chronic bacterial infections worldwide \cite{camara2022economic}. However, biofilms are considered to be vital in environmental remediation processes, resource recovery from wastewater, microbe-catalyzed electrochemical systems for energy, and the  production of commercially valuable products through genetic engineering or being stimulated with molecules to trigger a response \cite{mukhi2022beneficial,camara2022economic}.

Bacteria can chemically communicate cell-to-cell via extracellular signalling molecules called autoinducers (AIs). This process is called quorum sensing (QS) and relies on the production of and response to the AIs. Neighboring communities of bacteria have the ability to synchronously alter their behaviour in response to QS from changes in population density and species composition \cite{mukherjee2019bacterial}.  For example, the  production of green fluorescent protein (GFP) in the well-studied species \(\textit{Vibrio fischeri}\) occurs in the presence of the AI N-acyl homoserine lactone (AHL). Each individual cell behaves as both a transmitter (TX) and a receiver (RX) of AHL. Once there are sufficient cells releasing AHL, such that a threshold AHL concentration is observed around individual cells, the GFP genes switch on and fluorescence can be observed \cite{bai2015performance}. 

Furthermore, spatial heterogeneities in biofilms, such as variations in cell size and mass, are common and affect internal nutrient uptake \cite{wimpenny2000heterogeneity}. This is especially true in mature biofilms, where variations can lead to local nutrient deprivation and deplete growth. Therefore, multiple transport mechanisms are required for nutrients to reach all parts of a biofilm \cite{quan2022water}. 

The primary transport mechanism in biofilms is diffusion. Diffusion via the bacterial EPS alone might be too slow for molecules to be readily distributed throughout the whole biofilm within a sufficient timescale, as the EPS density contributes to a lower diffusion rate. Fortunately, soluble signalling molecules, nutrients, and waste can also be transported throughout biofilms via water channels. Water channels provide less obstructed pathways, such that molecules can propagate much more efficiently \cite{wilking2013liquid ,rooney2020intra}. Typical minimum dimensions for a water channel are a length of 1\,$\mu$m and a width of 100\,nm \cite{quan2022water}.

In mature biofilms, the role of bacteria at specific locations becomes more specialized as distinct genotypes are expressed in different regions, thereby establishing more defined roles within the ecosystem \cite{nesse2018biofilm}. Compared to planktonic (i.e., freely suspended) cells, biofilm-associated cells exhibit gene expression profiles indicating the loss of flagella, the development of antibiotic resistance mechanisms, and the production of biofilm matrix components \cite{rumbaugh2020biofilm}. The development of a defined architecture and heterogeneous biofilm structure is attributed to the differential regulation of certain genes by bacteria at the biofilm boundary. This regulation occurs in response to stimuli such as intercellular signaling, nutrient availability, and environmental gradients, resulting in unique genotypes at the boundary compared to those found at the center \cite{donlan2002biofilms}. Consequently, the gene expression patterns observed across the biofilm reflect adaptations to local environments, providing the stability necessary for the biofilm to thrive. 

Whether we are interested in the enhancement or disruption of biofilm  activity, a greater understanding of the local propagation characteristics of QS within biofilms will support our development of strategies to do so. In this direction, there has been significant interest in the mathematical modeling of QS systems with the support of simulations \cite{kannan2018mathematical,perez2016mathematical,li2020quantitative}. These models are diverse, with various approaches examining the production and transport of AIs, as well as the dynamics of cellular growth and division. One method to differentiate between models is to divide
them into deterministic and stochastic modelling. For further information, refer to \cite{perez2016mathematical,li2020quantitative}.

Molecular communication (MC) is a sub-field in communication engineering in which information is encoded using natural or synthetic molecules \cite{akan2016fundamentals}. Research in  MC can provide a greater understanding of the propagation of AIs. By understanding how molecular signals are transmitted and propagate within biofilms, we can develop strategies that improve the effectiveness of inhibition of antimicrobial agents. With recent advancements in QS disruptors, targeting QS is seen as a promising strategy for disrupting biofilm formation. This approach shows significant potential for inhibiting bacterial communication \cite{bjarnsholt2007quorum,paluch2020prevention}. Additionally, targeting QS is utilized in the bioprocessing industry to enhance the efficiency and yield of valuable products \cite{rosero2021microbial}. Predominantly, most MC models are based on uniform (isotropic) diffusion. However, the presence of the water channel network suggests that conventional isotropic diffusion is inappropriate for modeling signal propagation within a biofilm, such that anisotropic (i.e., \textit{non-uniform}) diffusion, where the diffusion coefficient varies with direction, should be considered, as proposed in \cite{van2012anisotropic}.

In MC, bacterial communication has been widely studied and well-established. A non-exhaustive list of topics includes bacterial cooperation \cite{noel2017effect}, bacterial biosensors \cite{martins2022microfluidic}, relaying \cite{einolghozati2013relaying}, machine learning applied to calcium signalling\cite{balasubramaniam2023realizing}, delay \cite{cobo2010bacteria}, and genetic circuits \cite{unluturk2015genetically} in idealized scenarios. QS in MC has been studied  with the aim of targeting biofilm disruption. Methods for modeling biofilm disruption via QS include signal jamming \cite{martins2018molecular}, queuing models \cite{michelusi2016queuing}, starvation-induced disruption \cite{martins2016using}, and QS mimickers \cite{gulec2023stochastic}. More generally, the theoretical limits of QS-based communication have been studied by analyzing channel capacity and bit-error rate. To simulate bacterial population dynamics, the stochastic birth-death processes are incorporated into the system \cite{tissera2020bio}.There has been prior research on molecular diffusion channels between bacterial clusters, but it did not account for bacteria behavioral responses \cite{einolghozati2013design}. Another study utilized stochastic geometry and probability processes to predict bacteria cooperation in a 2D environment \cite{fang2020characterization}. QS-based synchronization in nano-networks has been employed to amplify molecular communication signals \cite{abadal2012quorum}, explore the relationship between bacterial density and AI concentration \cite{tissera2019quorum}, and analyze how spatial dimensions, concentration thresholds, and inter-node distances influence oscillation periods and phase synchronization \cite{li2015evaluation}. In MC, work on bacteria and bacteria in biofilm-based communication has been carried out. However, despite the aforementioned development of MC models for biofilms, anisotropic diffusion has not been examined, except in our previous preliminary work \cite{paramalingam2023biofilm}.

There are many cases where anisotropic diffusion is present in biological systems \cite{li2024anisotropic}. The use of specific diffusion tensors in both 2D and 3D environments can be applied to anisotropic diffusion. 3D models were developed to aid in the imaging of biological tissue \cite{hanyga2014new}, but 2D models were used to study neurodegenerative disease progression through protein diffusion \cite{kevrekidis2020anisotropic}. Anisotropic diffusion
has diverse applications in 2D and 3D systems. For example, it is used to study deformable media and the interactions between cellular processes and mechanical stress, such as in cardiac muscle \cite{cherubini2017note}. In 3D, anisotropy describes molecular diffusion during eukaryotic cell division \cite{pawar2014anisotropic}, in skeletal muscles \cite{kinsey1999diffusional}, in extracellular vesicles in the cardiac matrix \cite{rudsari2022end}, during the motion of torqued swimmers \cite{sandoval2013anisotropic}, in brain tissue \cite{al2019influence}, and during antibody binding kinetics in drug delivery \cite{chahibi2015molecular}. A 3D anisotropic diffusion model based on microscopy images of single-species biofilms has shown that diffusion is anisotropic and depth dependent \cite{van2012anisotropic}. Although anisotropic diffusion in biofilms has been previously studied, to the best of our knowledge, there has been no research investigating the positioning of  TX and RX in a 2D model or comparing the impact of anisotropic diffusion against isotropic diffusion on signal propagation in a bounded system.

The aim of this paper is to establish an anisotropic diffusive 2D  system model for a biofilm and show how anisotropic diffusion enables information to propagate from different locations within the bounded space. The main contributions of this paper are summarized as follows:

\begin{itemize}
    \item We propose a simple 2D biofilm communication system within a reflecting bounded environment, featuring a point-to-point TX and RX. We account for the anisotropic transport with water channels by defining distinct diffusion coefficients in the radial and azimuthal directions. 
    \item The corresponding Green’s function for concentration (GFC) is derived as the channel impulse response for this system.
    \item Our results are validated through particle-based simulations (PBS), confirming the accuracy and reliability of our model.
    \item We investigate the impact of different system parameters on the characterization of the channel, including biofilm size, TX and RX locations, and the diffusion coefficient values. We use the results to gain intuition about communication and transport within biofilms, e.g., how water channels accelerate signaling between the edge and center of a biofilm.
    
\end{itemize}

The rest of the paper is organized as follows. Section II describes the biofilm structure and the anisotropic diffusive MC system within the biofilm. In Section III, the system's GFC is derived. The results and discussions are presented in Section IV and finally, Section V concludes the paper.

\section{System model}

In this section, we  construct a point-to-point diffusion model to explain the propagation of AIs diffusing across a biofilm. Our system model will focus on communication within a mature biofilm under ideal conditions, such that the water channels are fixed within a 2D environment with a circular boundary informed by \cite{quan2022water,wilking2013liquid,rooney2020intra}. As mentioned previously, AIs not only diffuse along the water channel, but can also diffuse across the rest of the EPS. Thus, we can model a biofilm as a porous material where the diffusion is anisotropic. To describe the anisotropic diffusion of the AIs within the biofilm, we employ the polar coordinate system where (\(\rho\), \(\theta\)) denote radial and azimuthal coordinates, respectively. In this model, we utilize a passive/transparent receiver and the radius of the biofilm is denoted by \(\rho_c\). It is assumed that the AIs released into the environment will degrade, be consumed, or transform into another molecule at a certain lumped rate, \(k_d\) $s^{-1}$. Hence, we model this conversion as a first-order degradation reaction, i.e.,

\begin{equation}
    A \xrightarrow{k_d}\phi.
\end{equation}

The circular boundary is assumed to be fully reflective where the AI $(\textit{A})$ will collide with the boundary and be reflected. We model the collision via the reaction

\begin{equation}
    A \xrightarrow{k_f}A_{Bound},
\end{equation}

\noindent
where \( k_f \) is a constant. Since the boundary is reflective, \( k_f = 0 \) \(\text{m}^2 \cdot \text{s}^{-1}\).

\begin{figure}[t!]
\includegraphics[width=1\linewidth]{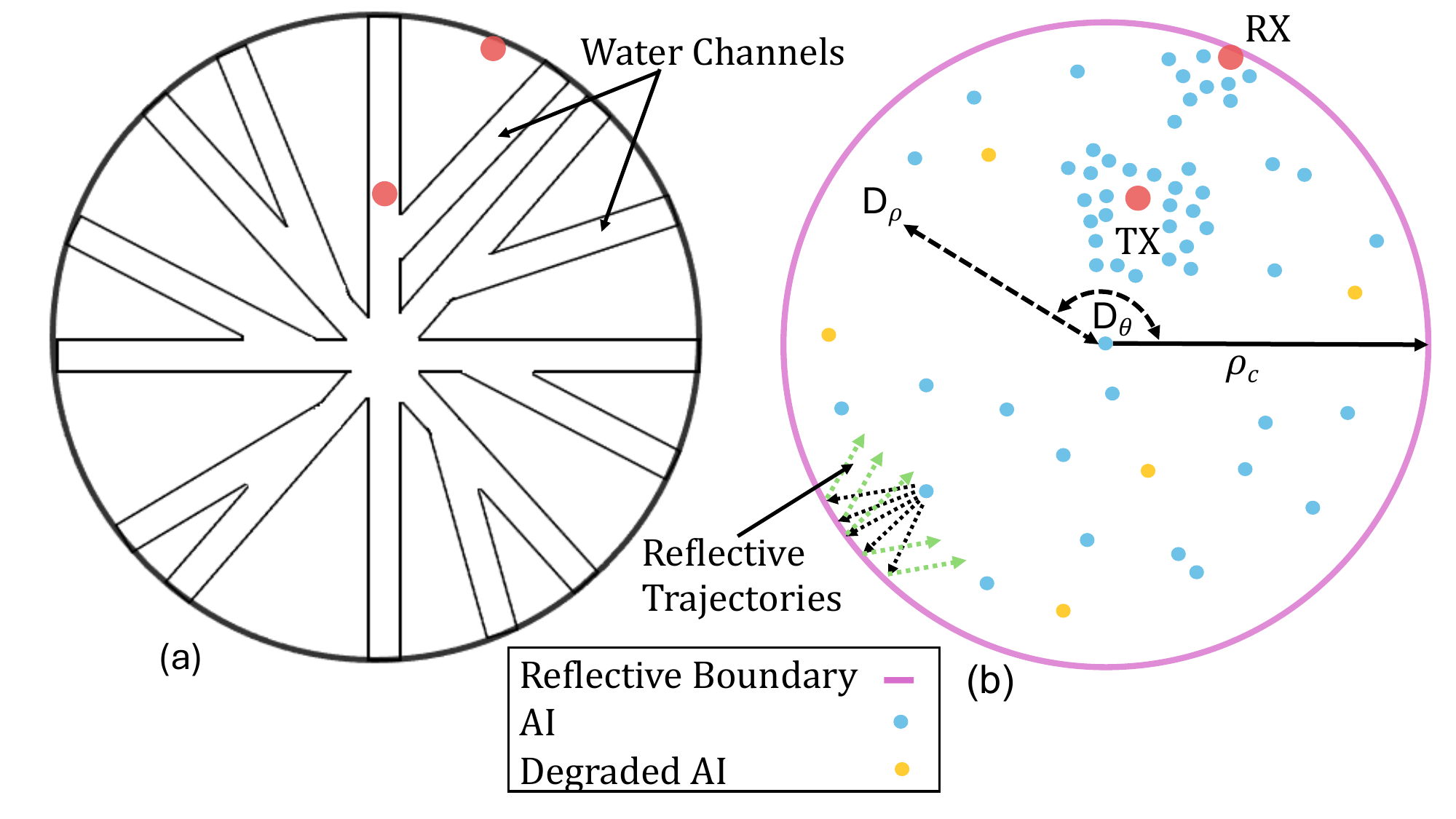}
\caption{Schematic representation of AI propagation in a 2D biofilm with point-to-point TX and RX. (a) Top-down view illustrating the water channels and the EPS. (b) An overview of the diffusion model detailing the propagation from TX to RX (passive/transparent), showcasing diffusion along $D_{\rho}$ and $D_{\theta}$ with a reflective boundary shown by multitude of arrows and degraded AI.}
\label{Figure 1}
\end{figure}

The microscopy images presented in \cite{wilking2013liquid} show that the water channels branch off the center and align in a manner forming a network, as represented in Fig.~\ref{Figure 1}(a). The radial water channels support the transmission of AIs in this direction to and from the edges of the biofilm, leading to directed diffusion along the radial axis. 
Thus, we represent the system in the simplified manner shown in Fig.~\ref{Figure 1}(b). We model the effective diffusion coefficient in the radial direction as $D_{\rho}(\rho)$, which varies with $\rho$ and is invariant to $\theta$, due to the symmetry. Considering there are no water channels aligned along the azimuthal direction, the diffusion in the azimuthal direction is expected to be slower than in the radial direction. Thus, we define the effective diffusion coefficient in the azimuthal direction as $D_{\theta}(\rho)$. $D_{\theta}(\rho)$ is invariant with respect to $\theta$ due to azimuthal symmetry. The location of the point source is assumed to be at $\Bar{r}\textsubscript{tx}$ = ($\rho\textsubscript{tx}$, $\theta\textsubscript{tx}$),
where 0 $\leq \rho\textsubscript{tx} \leq \rho_{c}$. The source has an instantaneous molecule release rate  $\delta(t-t\textsubscript{0})$  mol $\cdot$ $s\textsuperscript{-1}$, where $\delta$ $(\cdot)$ is the Dirac delta function and $t\textsubscript{0}$ is the release instant. $C(\bar r,t|{\bar r_{\mathrm{ tx}}},{t_{0}})$
denotes the concentration of molecules at point $\Bar{r}$ and  time ${t}$ giving the impulse point source $S(\Bar{r},t,\Bar{r}\textsubscript{tx},t\textsubscript{0}) = \delta\frac{(\rho-\rho\textsubscript{tx})}{\rho}\delta(\theta-\theta\textsubscript{tx})\delta(t-t\textsubscript{0})$. The diffusion of the autoinducer within the biofilm will be governed by the following partial differential equation with the boundary condition $\frac{\partial C}{\partial \rho} = 0$

\begin{multline}\label{1}
\frac {{\partial C(\bar r,t|{\bar r_{\mathrm{ tx}}},{t_{0}})}}{\partial t} = \mathbf{\nabla} \cdot \biggl(D\textsubscript{eff}\cdot \Vec{\mathbf \nabla}\ C(\bar r,t|{\bar r_{\mathrm{ tx}}},{t_{0}})
\\
- k\textsubscript{d}C(\bar r,t|{\bar r_{\mathrm{ tx}}},{t_{0}}) + S(\bar r,t|{\bar r_{\mathrm{ tx}}},{t_{0}})\biggr),
\end{multline}

where
\begin{equation}
     D\textsubscript{eff} = \begin{bmatrix} D_{\rho}  & 0 \\0 &  D_{\theta}  \end{bmatrix}.
\end{equation} \\
Solving (3) and (4) to find $C(\bar r,t|{\bar r_{\mathrm{ tx}}},{t_{0}})$ will enable us to explore how information propagates from different locations within a bounded biofilm.

\section{GFC of Diffusion in a Circle}
This section details the theoretical analysis of a QS system model within a 2D biofilm. We employ a diffusion-based MC model to explain the propagation of AIs. To analyze the diffusion of AIs in this 2D system, we model the channel impulse response from a point-to-point transmission from TX to RX. This analysis incorporates the effective diffusion tensor  $D\textsubscript{eff}$ in both the radial $\rho$ and angular $\theta$ directions. Thus, the GFC is formulated to characterize the expected diffusion signal.

\subsection{GFC Problem Definition}
To align  with the anisotropic properties of a biofilm as described in \cite{van2012anisotropic}, we model the biofilm as a porous medium. The third type (Robin) boundary condition of \cite{deng2015modeling} is used to characterize the reflection of the AI at the  boundary described in equation (2). Taking the divergence over the gradient of the AI concentration with the diffusivity tensor $D\textsubscript{eff}$ in $\rho$ and $\theta$ directions provides the concentration dynamics due to diffusion, thus (3) can be re-rwitten as 

\begin{equation*}
    D_{\rho}\frac {{\partial ^{2}{C_{\rho \theta }}(\rho,\theta,t|{\bar r_{\mathrm{ tx}}},{t_{0}})}}{{\partial {\rho ^{2}}}}+ \frac {{D_{\rho}}}{\rho }\frac {{\partial {C_{\rho \theta }}(\rho,\theta,t|{\bar r_{\mathrm{ tx}}},{t_{0}})}}{\partial \rho } + 
\end{equation*}

\begin{equation*}
   \frac {D_{\theta}}{\rho ^{2}}\frac {{\partial ^{2}{C_{\rho \theta }}(\rho,\theta,t|{\bar r_{\mathrm{ tx}}},{t_{0}})}}{{\partial {\theta ^{2}}}} - k_{d}C(\bar r,t|{\bar r_{\mathrm{ tx}}},{t_{0}}) + 
\end{equation*}

\begin{equation}
  \frac {{\delta (\rho - {\rho _{\mathrm{ tx}}})}}{\rho }\delta (\theta - {\theta _{\mathrm{ tx}}})\delta (t - {t_{0}})  =\frac {{\partial {C_{\rho \theta }}(\rho,\theta,t|{\bar r_{\mathrm{ tx}}},{t_{0}})}}{\partial t},
\end{equation}

\noindent
subject to the boundary condition 

\begin{equation}
D_{\rho}\frac {\partial C(\bar r,t|{\bar r_{\mathrm{ tx}}},{t_{0}})}{\partial \rho }\Bigg| _{\bar r=(\rho _{c},\theta)}=-k_{f} C(\rho _{c},\theta,t|{\bar r_{\mathrm{ tx}}},{t_{0}}),
\end{equation}

\noindent
and the GFC of diffusion is $C(\bar r,t|{\bar r_{\mathrm{ tx}}},{t_{0}})$. The GFC can be expressed as the product of a 1D and a 2D Green's function, as stated in \cite{cole2010heat}, i.e.,

\begin{equation} {C(\bar r,t|{\bar r_{\mathrm{ tx}}},{t_{0}})}=C_{\rho \theta }(\rho,\theta,t|\rho _{\mathrm{ tx}},\theta _{\mathrm{ tx}},{t_{0}}).\end{equation}

Applying (7) to boundary condition (6) results in

\begin{equation*}
   D_{\rho}\frac {\partial {C_{\rho \theta }}(\rho,\theta,t|{\rho _{\mathrm{ tx}}},\theta _{\mathrm{ tx}},{t_{0}})}{\partial \rho }\Bigg|_{\rho = \rho_{c}}
\end{equation*}

\begin{equation}
    =-k_{f} {C_{\rho \theta }}(\rho _{c},\theta,t|{\rho _{\mathrm{ tx}}},\theta _{\mathrm{ tx}},{t_{0}}). 
\end{equation}

Therefore, $C_{\rho \theta}(\rho, \theta, t | \rho_{\mathrm{ tx}}, \theta_{\mathrm{tx}}, t_0)$ is the solution of PDE (5) subject to the boundary condition (8).

\subsection{Derivation of Radial-Azimuthal GFC}

In this subsection, we solve \( (5) \) subject to the boundary condition \( (8) \). Our methodology is similar to the approach used in \cite{zoofaghari2018diffusive} to find the GFC for a cylindrical system  with  isotropic diffusion. The source term of $\delta\frac{(\rho-\rho\textsubscript{tx})}{\rho}\delta(\theta-\theta\textsubscript{tx})\delta(t-t\textsubscript{0})$ in \( (8) \) is equivalent to having the initial condition 

\begin{equation} C_{\rho \theta }(\rho,\theta,t=t_{0}|{\bar r_{\mathrm{ tx}}},{t_{0}})=\frac {{\delta (\rho - {\rho _{\mathrm{ tx}}})}}{\rho }\delta (\theta - {\theta _{\mathrm{ tx}}}).\end{equation}

Considering the initial condition and eliminating the source term in (5), a homogeneous partial differential equation is formulated, which can be solved using the separation of variables technique \cite{cole2010heat}. By substituting $C_{\rho\theta}(\rho,\theta,t|\bar{r}_{tx},t_0)=R(\rho|\rho_{tx})\Theta(\theta|\theta_{tx})T(t|t_0)$ in (5), in  which $R, \Theta$, and $T$ are functions of the corresponding variables that need to be found. Thus, in (5) without the source term and in the boundary condition (8), and dividing both sides by $R(\rho|\rho_{tx})\Theta(\theta|\theta_{tx})T(t|t_0)$, followed by some straightforward manipulation, the resulting problem becomes solving

\begin{equation*}
    \frac {\rho ^{2}R''(\rho |\rho _{\mathrm{ tx}})}{R(\rho |\rho _{\mathrm{ tx}})}\!+\!\frac {\rho R'(\rho |\rho _{\mathrm{ tx}})}{R(\rho |\rho _{\mathrm{ tx}})}\!-\!\frac {\rho ^{2} T'(t|t_{0})}{D_{\rho}T(t|t0)}-  \frac{k_d\rho^2}{D_\rho} \! =
\end{equation*} 

\begin{equation}
-\frac {D_{\theta}\Theta ''(\theta |\theta _{\mathrm{ tx}})}{D_{\rho}\Theta (\theta |\theta _{\mathrm{ tx}})}\overset {(a)}{=}\alpha, \\{}    
\end{equation}

\noindent
subject to the boundary condition

\begin{equation} D_{\rho}R'(\rho |\rho _{\mathrm{ tx}})\mid _{\rho =\rho _{c}} = -{k_{f}}R(\rho _{c}|\rho _{\mathrm{ tx}}).\end{equation}

The equality labeled (a) involving the constant $\alpha$ can be established due to the separation of variables present at the left and right hand sides of the initial equality. The notations prime ($'$) and double prime ($''$) denote the first and second derivatives of the function with respect to its solitary variable, respectively. Consequently, we obtain the subsequent ordinary differential equation

\begin{equation}
\Theta ''(\theta | \theta _{\mathrm{tx}}) + \alpha \frac{D_{\rho}}{D_{\theta}} \Theta (\theta | \theta _{\mathrm{tx}}) = 0.
\end{equation}

Given that the concentration varies periodically (with  period $2\pi$) with respect to the $\theta$ variable and is symmetric with respect to $\theta = \theta_\text{tx}$, the permissible values of $\frac{D_{\theta}}{D_{\rho}} $ are $\alpha \frac{D_{\rho}}{D_{\theta}}= n^2$ for all $n \in \mathbb{Z}^+$, where $\mathbb{Z}^+$ represents non-negative integers. As a result, an acceptable solution for (12) is $\Theta_n(\theta | \theta_\text{tx}) = G_n \cos(n(\theta - \theta_\text{tx}))$, where $G_n$ is an unknown constant. Considering $ \alpha \frac{D_{\rho}}{D_{\theta}} =n^2$ in (10) and some simple manipulation, we rewrite (10) as

\begin{equation}
 \frac {D_{\rho}R''(\rho |\rho _{\mathrm{ tx}})}{R(\rho |\rho _{\mathrm{ tx}})}+\frac {D_{\rho} R'(\rho |\rho _{\mathrm{ tx}})}{\rho R(\rho |\rho _{\mathrm{ tx}})}-\frac {D_{\theta}n^2}{\rho ^{2}}=\frac {T'(t|t_{0})}{T(t|t_{0})}\overset {(b)}{=}-\gamma_{n}^{2}.  
\end{equation} 

In equality (b), the presence of a constant in the equation is a result of the separation of variables on both sides of the first equality. We note that only a negative constant on the right side is possible, since a non-negative constant leads to unbounded function $T(t|t_0)$ and a correspondingly unbounded concentration function of time, which is impossible. By defining $\lambda_n = \sqrt{\frac{\gamma_n}{D_\rho}}$ in  (13), we obtain the desired result

\begin{equation}
\rho ^{2} R_{n}''(\rho |\rho _{\mathrm{ tx}})+\rho R_{n}'(\rho |\rho _{\mathrm{ tx}})+ \left(\lambda _{n}^{2}\rho ^{2} - \alpha \frac{D_{\rho}}{D_{\theta}}\right)R_{n}(\rho |\rho _{\mathrm{ tx}})= 0,    
\end{equation}

\noindent
with the boundary condition

\begin{equation} {D_{\rho}} R_{n}'(\rho |\rho _{\mathrm{ tx}})\mid _{\rho =\rho _{c}} = -{k_{f}}R_{n}(\rho _{c}|\rho _{\mathrm{ tx}}).\end{equation}

Eq. (14) is Bessel’s equation with general solution \cite{cole2010heat}

\begin{equation}
R_{n}(\rho|\rho _{\mathrm{ tx}}) = A_{n}J_{\zeta}(\lambda_n\rho) + B_{n}Y_{\zeta}(\lambda_n\rho),
\end{equation}

\noindent
where $\zeta=\sqrt\frac{D_{\theta}}{D_{\rho}}n$, $A_n$ and $B_n$ are unknown constants, and $J_\zeta(\cdot)$ and $Y_\zeta(\cdot)$ represent the Bessel functions of order $\zeta$ of the first and second kind, respectively, for every positive value $\lambda^2_n$. Since $Y_\zeta = \left(\sqrt{\frac{D_{\theta}}{D_{\rho}}}n\right)$ is singular at $\rho = 0$, we set $B_n = 0$. Substituting $R_n(\rho|\rho_\text{tx}) = A_{n}J_{\zeta}(\lambda_n\rho)$ in the boundary condition (15), we need to satisfy
\begin{equation} 
D_{\rho}{\lambda _{n}}{J_{\zeta}}'{\zeta}A_{n}J_{\zeta}(\lambda_n\rho) =-k_{f}J_{\zeta}A_{n}J_{\zeta}(\lambda_n\rho).
\end{equation}

In the context of the boundary condition (15), all roots of (17) except $\lambda_n=0$ for $n>0$ are acceptable $\lambda_n$ values, as $\lambda_n=0$ leads to the trivial solution of $R_n(\rho|\rho_{tx})=0$. It should be noted that $\lambda_0=0$ is a root for the boundary condition (17) when $k_f=0$, which results in the solution $R_0(\rho|\rho_{tx})=A_0$. If we denote the ${m}$th root of the aforementioned equation as $\lambda_{nm}$, then the solution to (14) with boundary condition (15) is given by $R_{nm}(\rho_c|\rho_{tx})=A_{nm}J_\zeta(\lambda_{nm}\rho) $. To satisfy equation (13) and considering the implicit condition of $\lim_{t\rightarrow\infty}T(t|t_0)=0$, we can use $T(t|t_0)=I_{nm}e^{(-D_\rho \sqrt{\lambda_{nm}})(t-t_0)}u(t-t_0)$, where $I_{nm}$ is an unknown constant and $u(\cdot)$ represents the step function. Thus, the solution to (5) is of the general form

\begin{equation*}
{C_{\rho \theta }}(\rho,\theta,t|{\bar r_{\mathrm{ tx}}},{t_{0}}) = 
\sum\limits_{n=0}^\infty\sum\limits_{m=1}^\infty (H_{nm}J_{\zeta}(\lambda_n\rho) \\
\end{equation*}
\begin{equation}   
\rho \cos(n(\theta-\theta_{\mathrm{ tx}})) e^{-D_\rho \sqrt{\lambda_{nm}}(t-t_{0})}u(t-t_{0})).
\end{equation}

The equation $H_{nm}=G_nA_{nm}I_{nm}$ is not known and needs to be solved using the initial condition provided in (9). The expansions of the delta functions $\delta(\theta-\theta_{\mathrm{tx}})$ and $\delta(\rho-\rho_{\mathrm{tx}})\rho$ are given in \cite{duffy2015green} as

\begin{equation} \delta (\theta - {\theta _{\mathrm{ tx}}}) =\sum \limits _{n=0}^\infty {L_{n} \cos (n(\theta - \theta _{\mathrm{ tx}}))},\end{equation}

\noindent
where

\begin{equation*}
 L_{0}=\frac{1}{2\pi}, \quad L_{n}=\frac{1}{\pi}, 
 n\geq 1,\text{ and }
 \end{equation*}

\begin{align}
\frac{\delta (\rho - \rho_{\mathrm{tx}})}{\rho} &= \sum_{m=1}^{\infty} \frac{J_{\zeta}(\lambda_n \rho)}{N_{nm}} J_{\zeta}(\lambda_n \rho), \\
N_{nm} &= \int_{0}^{\rho_c} \rho J_\zeta^2 (\lambda_n \rho) \, d\rho \nonumber \\
&= \frac{\rho_c^2}{2} (\lambda_n \rho_c) \nonumber 
 - J_{\zeta-1}(\lambda_n \rho_c) J_{\zeta+1}(\lambda_n \rho_c).
\end{align}

By applying (18)–(20) to initial condition (9) and comparing left and right sides of the equation, we obtain

\begin{equation}
H_{nm} = \frac{J_\zeta(\lambda_n\rho_c) L_{n}}{N_{nm}}, \quad n \geq 0,~ m \geq 1.
\end{equation}

Substituting (18) into (7), the GFC of diffusion in a 2D biofilm is obtained as

\begin{align}
C(\bar r,t) = & \sum_{n=0}^\infty\sum_{m=1}^\infty \biggl({\frac{{L_{n}{J_{\zeta}}\left(\lambda_n\rho\right)\rho\textsubscript{tx}}}{{{N_{nm}}}}} \nonumber \\
& \times{J_{\zeta}}\left(\lambda_n\rho\right)\rho\cos(n(\theta-\theta_{tx})) \nonumber \\
& \times e^{-D_\rho \sqrt{\lambda_{nm}}(t - t_{0})}u(t - t_{0})\biggr).
\end{align}

\section{Simulation and Numerical Results}

In this section, we investigate the influence of various system parameters on the GFC in molecular diffusion processes within a biofilm. A point-to-point molecular diffusion communication system is employed to evaluate performance efficacy. Evaluating the GFC computes the expected channel impulse response as a received concentration for the corresponding system. For numerical tractability of the GFC, the infinite sum in Eq. (23) is evaluated for \(n < 3\) and \(m < 5\); including additional eigenvalues ${\lambda_{nm}}$ had negligible impact on calculation accuracy. Unless otherwise specified, the system parameters considered are as listed in Table 1.

The numerical computation of the GFC is validated with a particle-based simulator (PBS) implemented in MATLAB (R2023a; the MathWorks, Natick, MA, USA). The PBS tracks molecule positions in 2D using polar coordinates and simulates their diffusion as independent random events. We approximate a point RX as a small circle in the PBS that we use to calculate an observed concentration and  $\theta$ = 0 rad for all instances. The movement of each molecule is simulated over discrete time intervals of $\Delta t$ and we average results over 500 realizations. Additionally, molecule degradation is modeled in accordance with (1), where each molecule has a probability of $e^{-k_d \Delta t}$ of degrading during any given interval, leading to its removal from the simulation. The system considered throughout this section has AIs diffusing within a circular 2D biofilm with a reflective boundary. The TX is a point source that is located at either the center of the circle, near the circle boundary, or at a location in between. Both isotropic and anisotropic diffusion scenarios are examined to reveal the features and benefits of anisotropic diffusion in biofilms.

\FloatBarrier
\begin{table}[t!]
\centering 
\caption{Parameters for the proposed biophysical model.} 
\label{tab:my_label} 
\resizebox{\columnwidth}{!}{
\begin{tabular}{@{}lr@{}} 
\toprule 
\textbf{Parameter} & \textbf{Value} \\
\midrule 
Diffusion coefficient in radial direction, \(D_{\rho}\) & \(5\times 10^{-10} \, \text{m}^2 \cdot \text{s}^{-1}\) \\
Diffusion coefficient in azimuth direction, \(D_{\theta}\) & \(\{5\times 10^{-10}, 5\times 10^{-11}\} \, \text{m}^2 \cdot \text{s}^{-1}\) \\
Circle radius, \(\rho _{c}\) & 100 \(\mu\text{m}\) \\
Point source transmitter location, \(\bar{r}_{\mathrm{tx}}\) & \(\{20, 40, 60, 80\}\) \(\mu\text{m}\), 0 rad \\
Degradation reaction constant, \(k_d\) & 0.3 \(\text{s}^{-1}\) \\
Receiver radius & 1 \(\mu\text{m}\) \\
Number of transmitted molecules inside the biofilm & \(1 \times 10^{7}\) \\
Time step in PBS, \(\Delta t\) & \(10^{-2}\) s \\
Number of realizations in PBS & 500 \\
\bottomrule 
\end{tabular}
}
\end{table}
\FloatBarrier

\subsection{Point-to-Point Channel Validation}

In Figs. 2, 3, and 4, we present a comparison of the GFC as calculated using our derived analytical result in Eq. (23) and compared against PBS data. All figures show a close match between the analytical derivations and the PBS results. The diffusion coefficients for isotropic diffusion are set to \(D_{\rho}=D_{\theta}=5 \times 10^{-10} \, \text{m}^2 \cdot \text{s}^{-1}\), whereas for anisotropic diffusion we reduce \(D_{\theta}\) to \(5 \times 10^{-11} \, \text{m}^2 \cdot \text{s}^{-1}\).

\begin{figure}[bt]
\raggedleft
\includegraphics[height=0.52\linewidth]{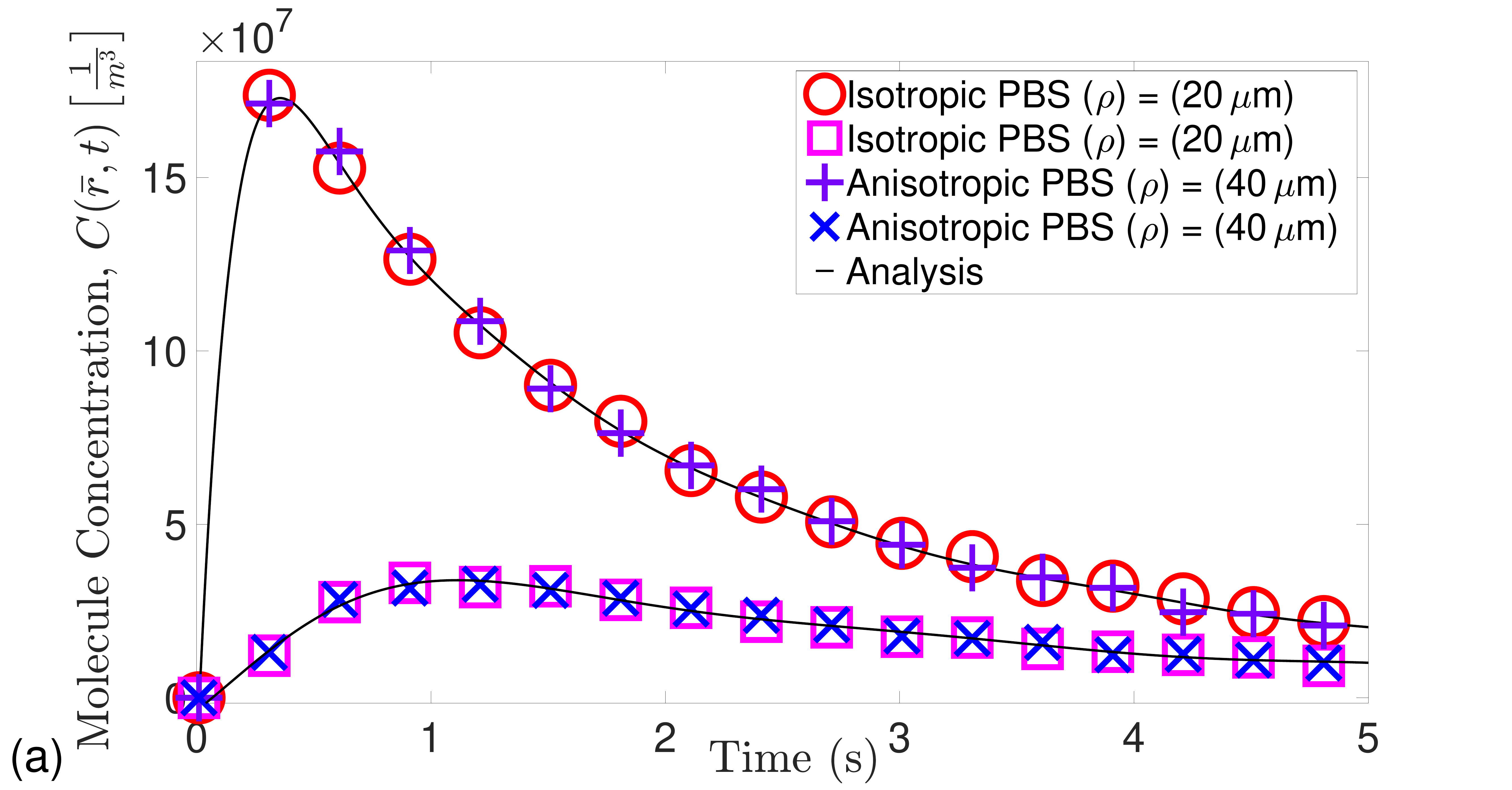}
\includegraphics[height=0.52\linewidth]{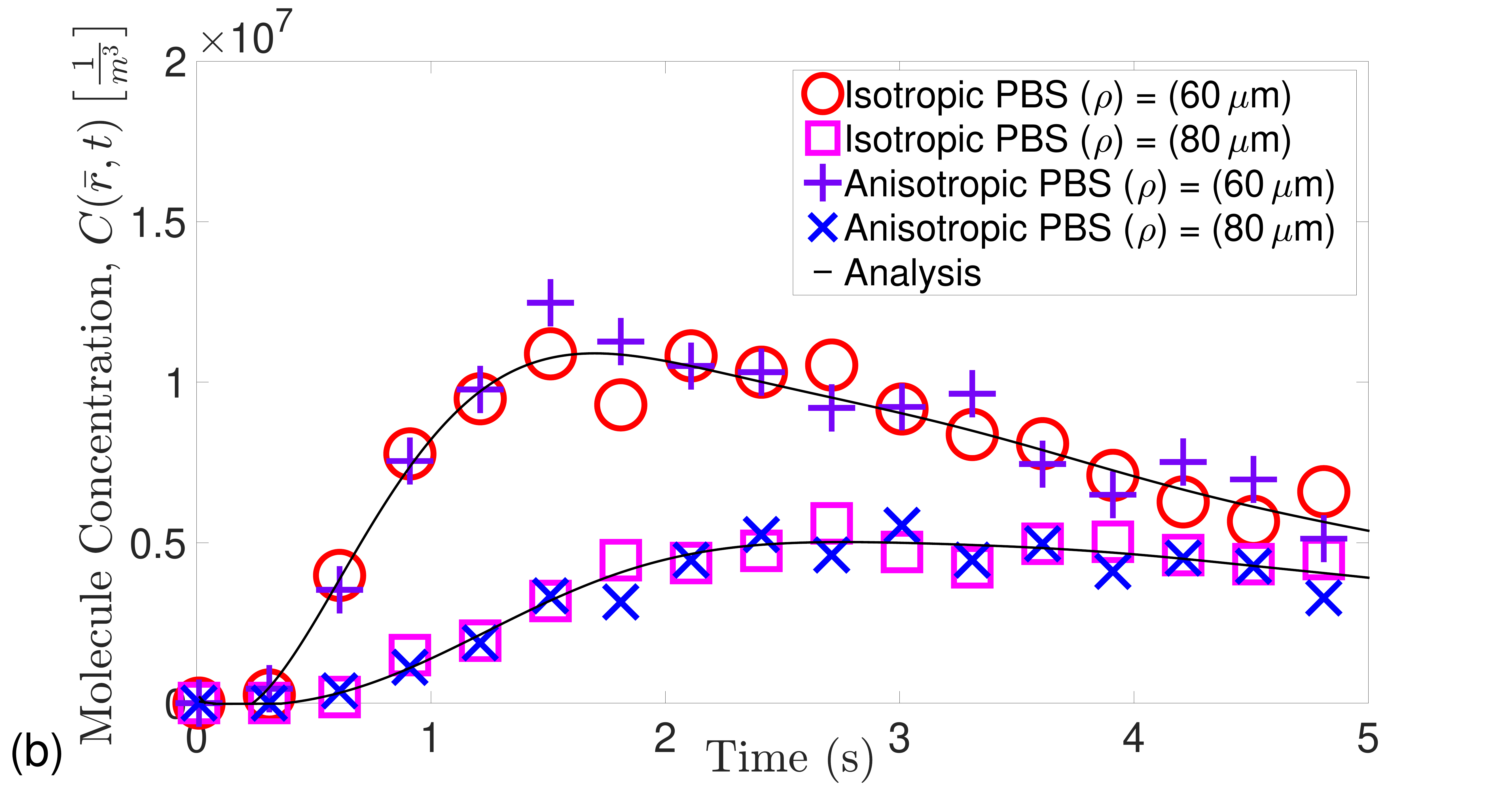}
\caption{Comparison of isotropic and anisotropic diffusion. The analytical concentration computed using (23) is compared with PBS. The transmitter location is at the biofilm center, represented as (0 $\mu$m, 0 rad). Observations were made at receiver points $\rho = \{20, 40\}$ $\mu$m in (a) and $\rho = \{60, 80\}$ $\mu$m in (b) with $\theta = 0$ rad, and $k_d = 0.3$ s\(^{-1}\).}
\label{Figure 2}
\end{figure}

Fig.~\ref{Figure 2}, to visualize the differences between isotropic and anisotropic diffusion profiles, the TX is centrally located with receivers placed at distances of $\rho$ = {\{20, 40, 60, 80\}} \(\mu\)m. No significant differences were observed between isotropic and anisotropic diffusion. The analytical calculations for both cases generated identical results, which was anticipated due to system symmetry when the TX is placed at the center. Additionally, the simulation results were consistent with the analytical calculations across all instances. As expected, the closest RX at 20 \(\mu\)m shows the highest peak concentration, while the furthest at 80 \(\mu\)m shows the lowest.

\begin{figure}[ht!]
\centering
\includegraphics[height=0.52\linewidth]{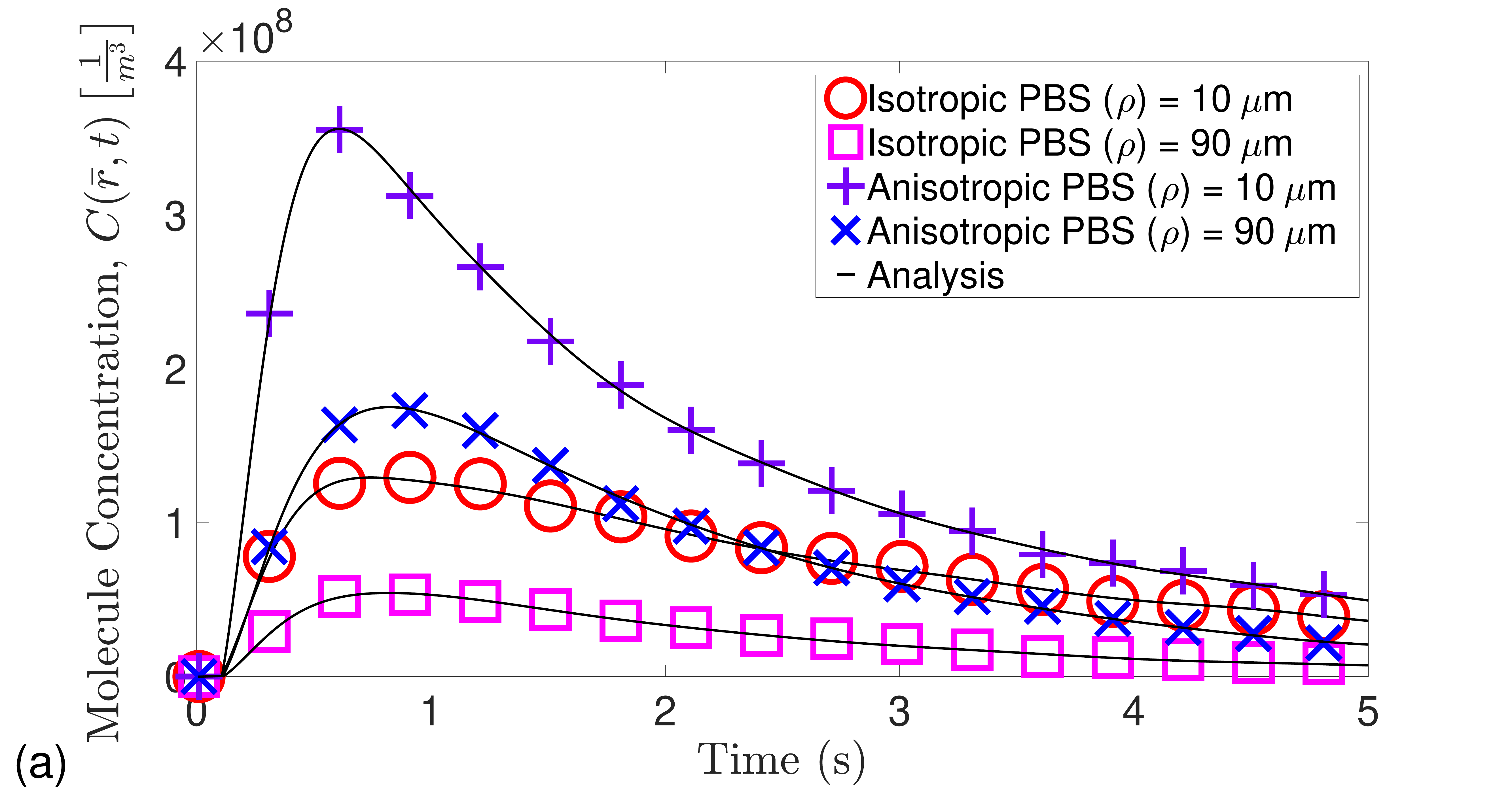}
\hspace{-15pt} 
\includegraphics[height=0.52\linewidth]{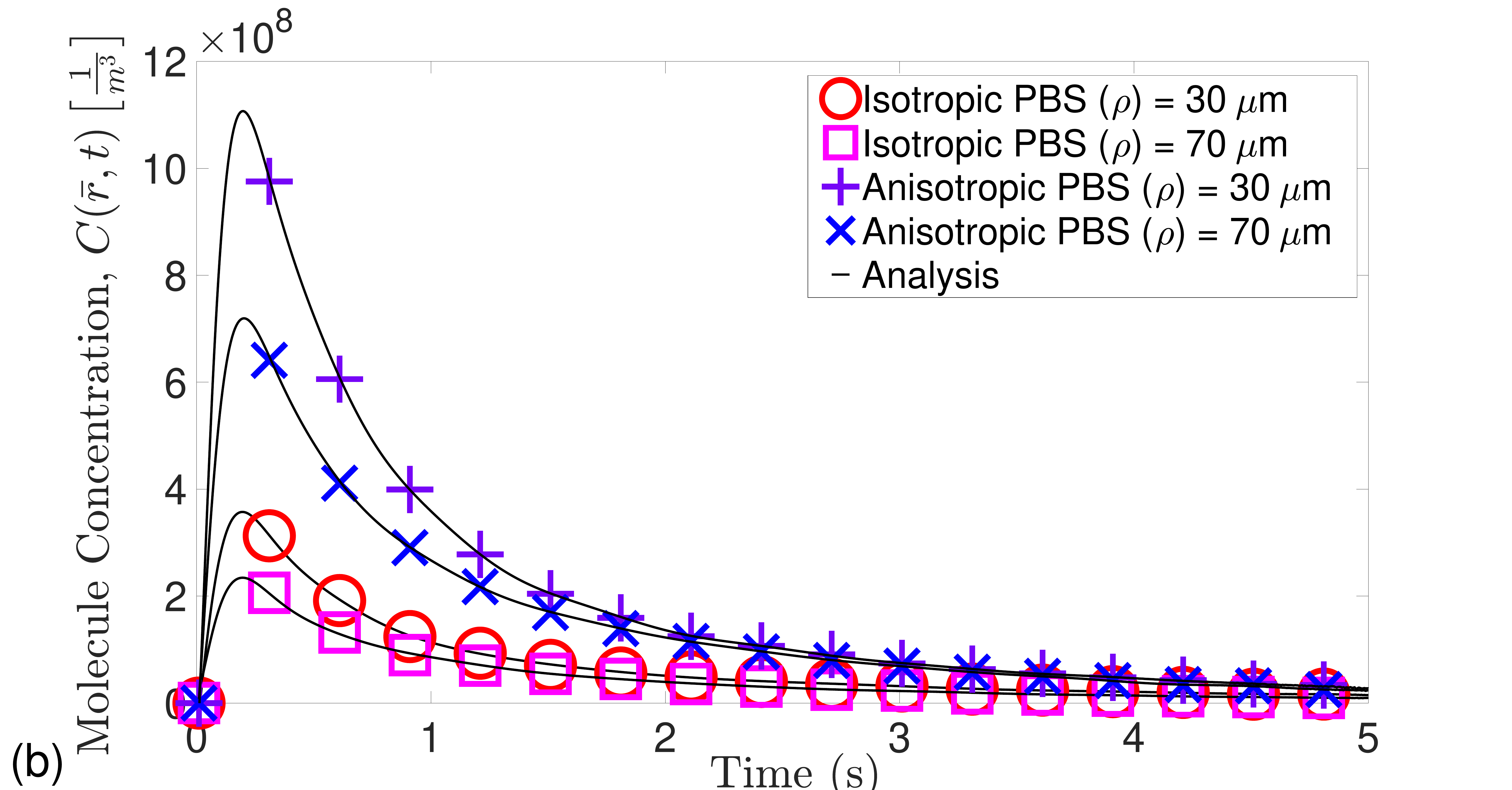}

\caption{Comparison of isotropic and anisotropic diffusion. The analytical concentration computed using (23) is compared with PBS. The TX is placed 50 $\mu$m from the center, represented as (50 $\mu$m, 0 rad). 
Observations were made at receiver points $\rho = \{10, 90\}$ $\mu$m in (a) and $\rho = \{30, 70\}$ $\mu$m in (b) with  $\theta = 0$ rad, and $k_d=0.3$ s\(^{-1}\).}
\label{Figure 3}
\end{figure}

In Fig.~\ref{Figure 3}, the effects of positioning the TX at 50 \(\mu\)m from the center were analyzed to compare the impacts of isotropic and anisotropic diffusion. All the simulation results were consistent with the analytical calculations. Anisotropic diffusion was observed to produce higher diffusion peaks compared to isotropic diffusion across all TX positions $\rho$ = \{10, 30, 70, 90\} \(\mu\)m from the center. Notably, the TX positioned at 30 \(\mu\)m from the center exhibited the highest diffusion peak, while the lowest peak occurred at 90 \(\mu\)m. The variation in concentration profiles at these distances from the TX can be attributed to the reflective boundary, which modifies the diffusion gradient. As a result of azimuthal diffusion, AIs near the boundary are displaced a greater net distance away when near the boundary.

\begin{figure}[ht]
\vspace{5pt} 

\includegraphics[height=0.52\linewidth]{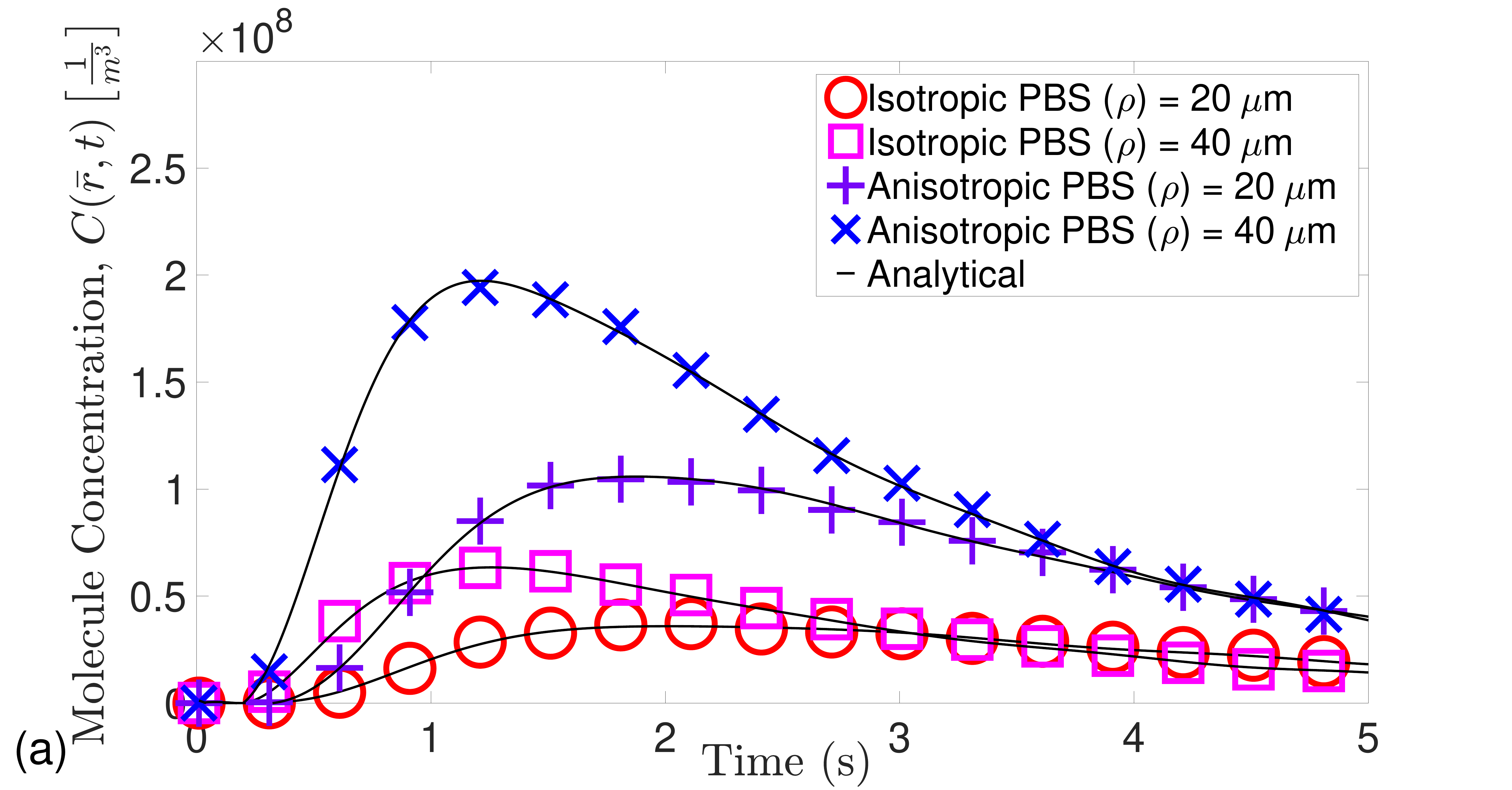}\\
\includegraphics[height=0.52\linewidth]{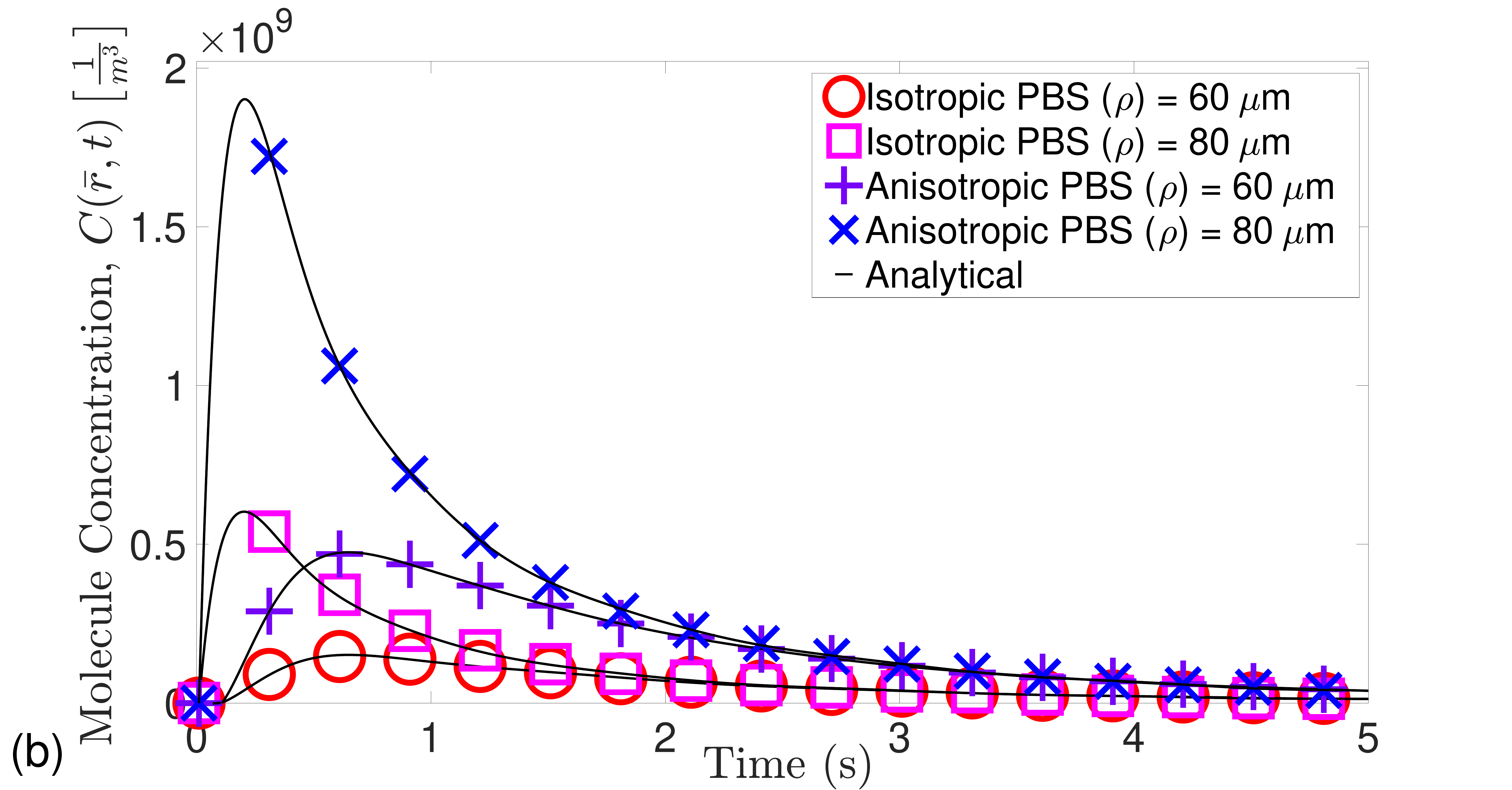}

\caption{Comparison of isotropic and anisotropic diffusion. The analytical concentration computed using (23) is compared with PBS. The TX located 100 $\mu$m from the center, represented as (100 $\mu$m, 0 rad). Observations were made at receiver points $\rho = \{20, 40\}$ $\mu$m in (a) and $\rho = \{60, 80\}$ $\mu$m in (b) with  $\theta = 0$ rad, and $k_d=0.3$ s\(^{-1}\).}
\label{Figure 4}
\end{figure}

In Fig.~\ref{Figure 4}, the RX locations are set at $\rho$ = \{20, 40, 60, 80\} \(\mu\)m from the center, while the TX is located at the boundary. The maximum diffusion peak was observed at  80 \(\mu\)m, as expected, while the minimum was noted at 20 \(\mu\)m. Similar to as observed in Fig.~\ref{Figure 3}, all RX positions have higher peak concentrations for anisotropic diffusion than for isotropic diffusion. Both Figs. 3 and 4 have demonstrated that anisotropic diffusion is more effective at propagating signals radially across the biofilm when the signal is not sent from the center of the biofilm (unlike Fig. 2, which shows the same results for isotropic and anisotropic diffusion). These observations align with our intuition, since we anticipated superior performance from anisotropic diffusion by generating higher diffusion peaks.

\subsection{Spatial Diffusion Profiles}
We further consider the spatial diffusion dynamics across the biofilm in Figs. 5, 6, and 7. To demonstrate the behaviour of different biofilm conditions, the following figures illustrate the AI concentrations at different times under varying diffusion coefficients and biofilm sizes. The color gradient represents 2D diffusion with the white outer circle indicating the boundary of the biofilm. Furthermore, a colorbar is present in all subfigures and transitions from yellow to blue, representing high to low concentrations, respectively. 
Additionally, Figs. 5-7 share the same pixel dimension of 20 \(\mu\)m.

\begin{figure}[ht]
\includegraphics[width=1\linewidth]{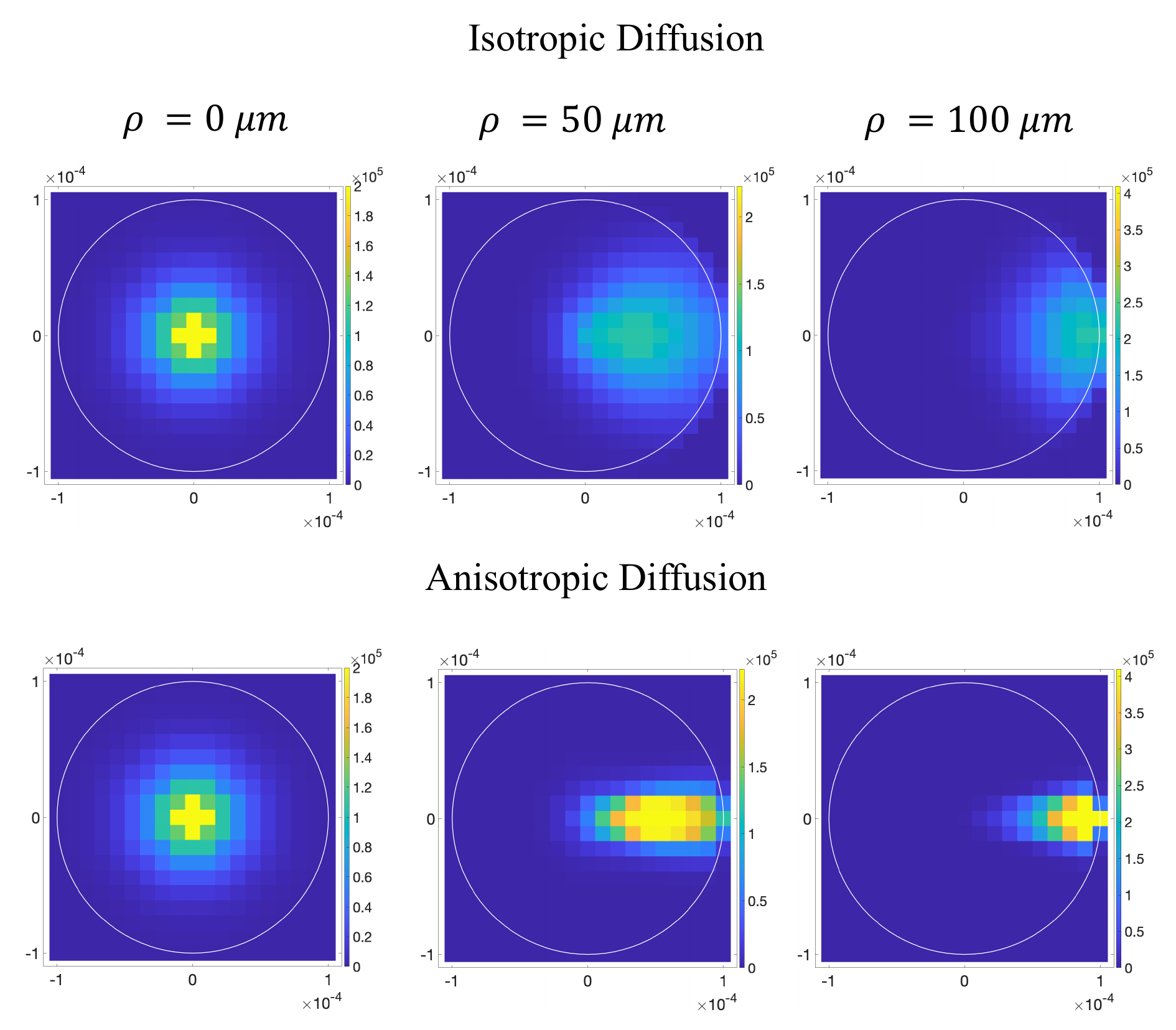}
\caption{2D diffusion profile of autoinducers from three different transmitter locations of $\rho$ = \{0, 50, 100\} \(\mu\)m at $\mathrm{t}$ = 3 s, where the the colorbar states the number of molecules. The top row is isotropic diffusion with $D_{\rho}$ = $D_{\theta} =5 \times 10^{-10}$ \(\text{m}^2 \cdot \text{s}^{-1}\) and the bottom row is anisotropic diffusion with $D_{\rho} = 5 \times 10^{-10}$ \(\text{m}^2 \cdot \text{s}^{-1}\) and $D_{\theta} = 5 \times 10^{-11}$ \(\text{m}^2 \cdot \text{s}^{-1}\). The white circle is the boundary of the biofilm and $k_d=0.3$ s\(^{-1}\).}
\label{Figure 5}
\end{figure}

In Fig. 5, both isotropic and anisotropic diffusion are presented. The top row displays isotropy, and the bottom row displays anisotropy. AIs are released from different radial positions of $\rho$ = \{0, 50, 100\} \(\mu\)m from the center, to compare the diffusion dynamics at positions similar to those in Figs. 2, 3, and 4. The general trend of isotropic diffusion is a more uniform diffusion of AIs in all placements of TX inside the biofilm. Similarly, as expected in Fig. 2, the spread of AIs across the biofilm is identical for both diffusion scenarios when $\rho = 0 \ \mu\text{m}$. Notably, the most drastic visual differences are pronounced at distances of 50 and 100 \(\mu\)m  from the center between isotropic and anisotropic diffusion, consistent with the results  in Figs. 3 and 4.

\begin{figure}[ht]
\includegraphics[width=1\linewidth]{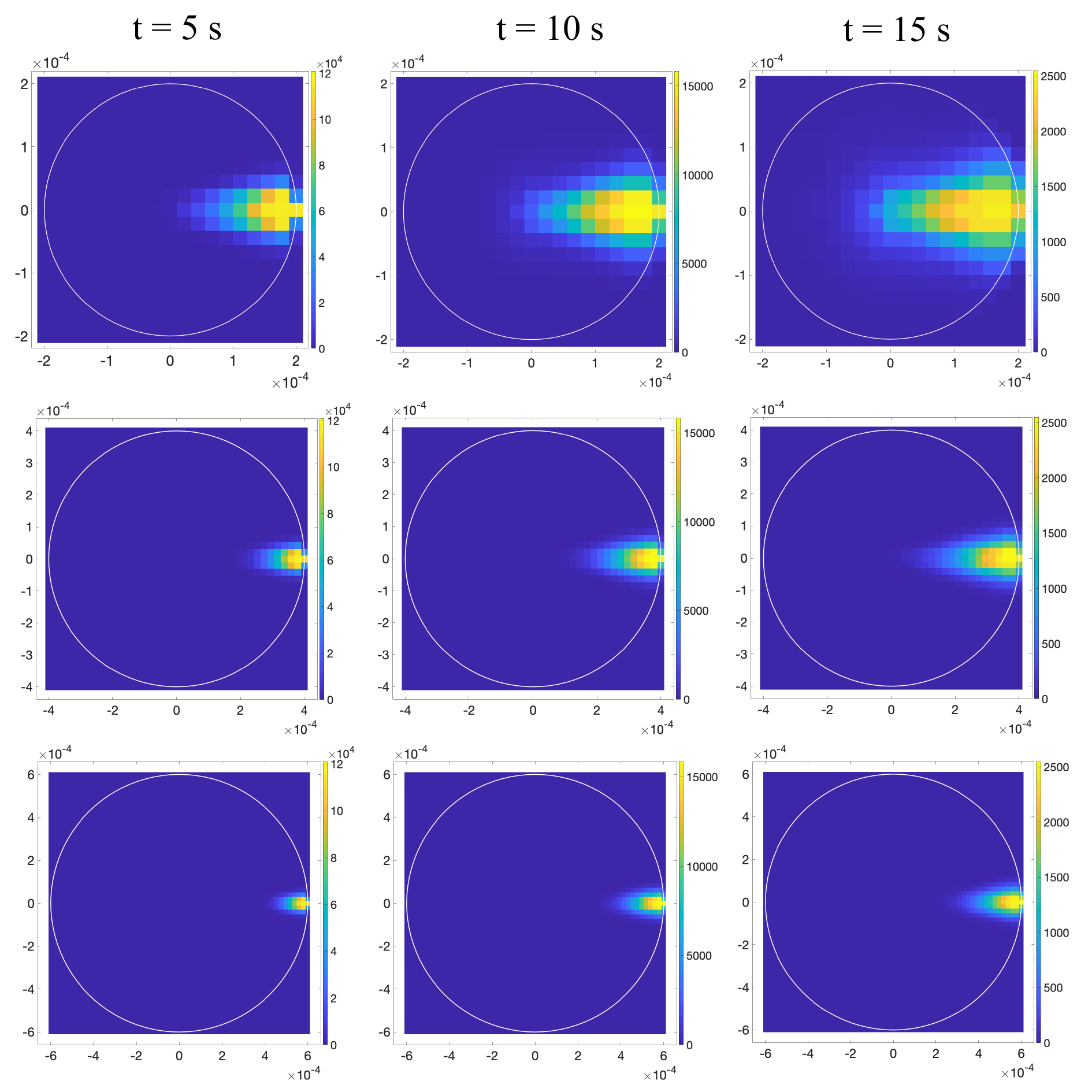}

\caption{2D diffusion profile consisting of  $D_{\rho} = 5 \times 10^{-10}$ \(\text{m}^2 \cdot \text{s}^{-1}\)and $D_{\theta} = 5 \times 10^{-11}$ \(\text{m}^2 \cdot \text{s}^{-1}\). The figure is organized into columns at $\mathrm{t}$ = \{5, 10, 15\} s and rows of different biofilm radii $\rho_{c}$ = \{200, 400, 600\} \(\mu\)m, respectively. The boundary is reflective and $k_d=0.3$ s\(^{-1}\).}
\label{Figure 6}
\end{figure}

In Fig. 6, colormaps are employed to demonstrate the spatiotemporal variations due to anisotropy when the TX is placed at the boundary. The figure is organized into columns and rows; the columns document temporal snapshots at $\mathrm{t}$ = \{5, 10, 15\} s, while the rows represent biofilm radii $\rho_{c}$ = \{200, 400, 600\} \(\mu\)m. Over time, molecules degrade at a rate of  \(k_d=0.3\) \(s^{-1}\). The most extensive diffusion across the biofilm is evident in the top row, since this is the smallest biofilm. Conversely, the least AI coverage is observed in the bottom row, with a radius of 600 \(\mu\)m. These results confirm that there is a smaller (relative) spread when the biofilm is larger, aided by water channels that enhance transport from the center to the boundary, thus highlighting the diffusion dynamics within the biofilm. These findings are consistent with the analytical results shown in Fig.~\ref{Figure 4}.

\begin{figure}[ht]
\includegraphics[width=1\linewidth]{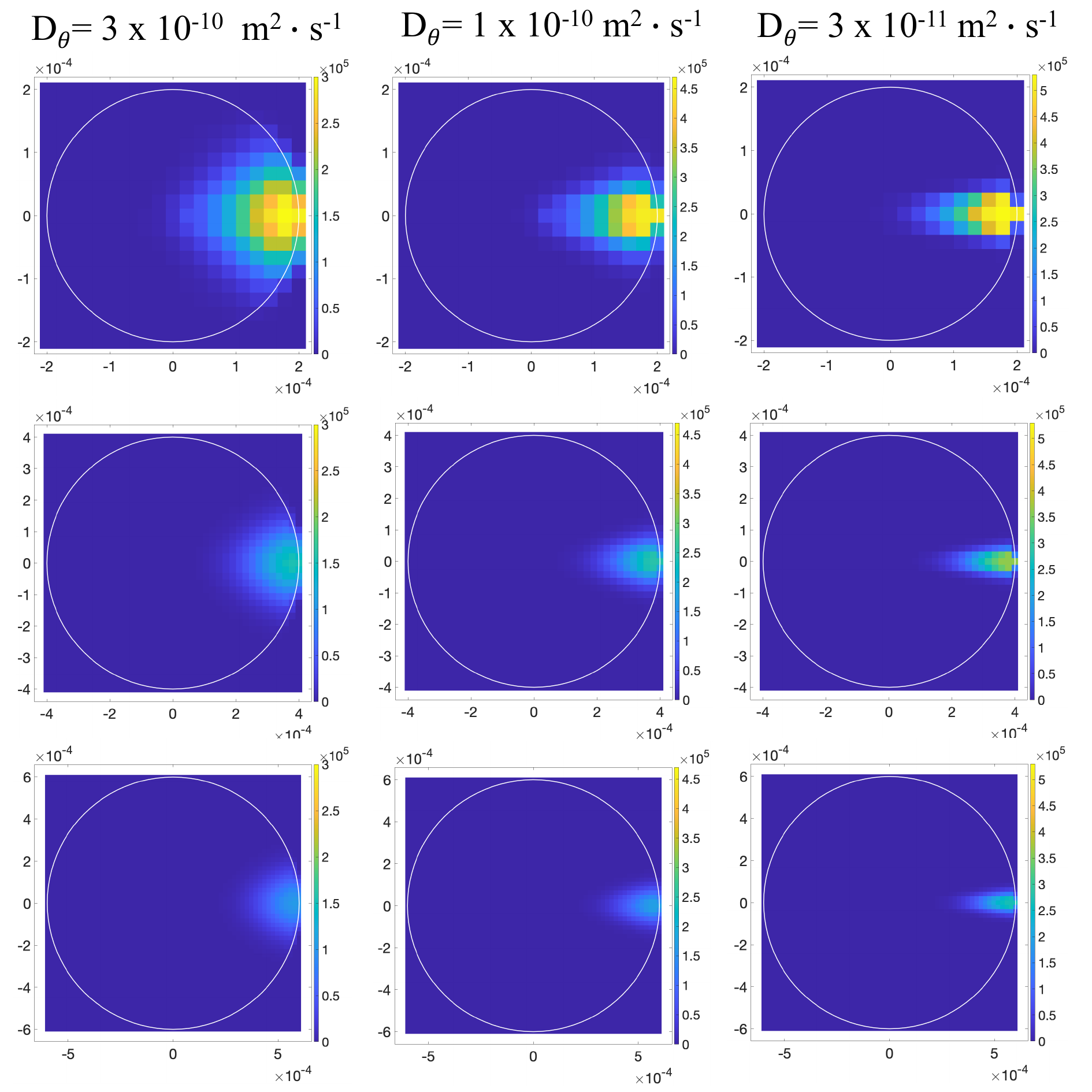}
\caption{2D diffusion profiles with three different levels of anisotropy. In all profiles, \(D_\rho = 5 \times 10^{-10}\) \(\text{m}^2 \cdot \text{s}^{-1}\)and \(D_\theta = \{3 \times 10^{-10}, 1 \times 10^{-10}, 3 \times 10^{-11}\}\) \(\text{m}^2 \cdot \text{s}^{-1}\) for the first, second, and third columns, respectively. Each row corresponds to biofilm radii \(\rho_{c} = \{200, 400, 600\}\) and \(k_d=0\) s\(^{-1}\).}
\label{Figure 7}
\end{figure}

In Fig. 7, detailed colormaps are presented that showcase anisotropic diffusion with the TX on the boundary under different levels of anisotropy. The figure demonstrates the spatiotemporal diffusion under different values of \(D_\theta = \{3 \times 10^{-10}, 1 \times 10^{-10}, 3 \times 10^{-11}\} \, \text{m}^2 \cdot \text{s}^{-1}\) from the first to the third column, respectively, while rows represent biofilm radii \(\rho_{c} = \{200, 400, 600\} \, \mu\text{m}\).
The figure reveals a greater degree of anisotropy as there is a narrower spread of AIs in the azimuthal direction with decreasing  \(D_\theta\). Under the limiting conditions of \(k_d\) = 0 and  $\mathrm{t}$ $\to \infty$ , the diffusion dynamics of the molecules remain consistent across different biofilm sizes, suggesting that the behavior of AIs does not vary with biofilm size under these conditions. Thus, regardless of the size and diffusion coefficients, the diffusion profile remains constant across all biofilms when \(k_d=0\) s\(^{-1}\).

\subsection{Discussion}

A  model that considers biofilms as a porous 2D medium and focuses on bacterial signal propagation has, to our knowledge, not been explored previously. Building on the understanding of the functioning of water channels \cite{wilking2013liquid} and anisotropic diffusion \cite{van2012anisotropic}, we have found that anisotropic diffusion plays a significant role within the biofilm. 

Our findings show that with a diffusion coefficient \(D_{\rho}\) ten times greater  than \(D_{\theta}\), when the TX is centrally located in our model, the diffusion profiles remain the same, as supported by Figs. 2 and 5 due to the symmetry. However, when the TX is off-center, anisotropic diffusion facilitates superior propagation profiles along the radial direction compared to isotropic diffusion, as depicted in Figs. 3, 4, 5, and 6. Hypothetically, this could function as a defense mechanism, enabling bacteria at the boundary layer to produce AIs that propagate more rapidly towards the core of the biofilm. From there, the signal can be transmitted back to the boundary, facilitating a two-way communication process.  

In pathogenic bacteria, quorum quenching enzymes are a set of antibacterial agents that inhibit key steps of quorum sensing (e.g., signal generation, accumulation, or reception) \cite{ilangovan2013structural}. With the rise of antibiotic-resistant bacteria, our model could be used to study the propagation of quorum-quenching enzymes  by placing them at the boundary and simulating the time required for them to reach the centre of the biofilm, resulting in biofilm penetration time. It is well established that molecular relaying occurs in biofilms. Therefore, incorporating relaying into our model in future studies would enhance the accuracy of our biofilm model.

\section{Conclusion}

In this paper, we proposed a point-to-point molecular communication (MC) system that introduces a 2D anisotropic diffusion model within biofilms, utilizing bacterial transmitters (TX) and receivers (RX). We derived the analytical solution for Green’s function for concentration (GFC), which serves as the channel impulse response, and validated it with PBS. To gain a deeper understanding of communication within biofilms, we characterized the channel by examining the impact of various system parameters including TX and RX locations, biofilm size, and the diffusion coefficient values. Our findings show that, due to symmetry, similar results occur under both isotropic and anisotropic diffusion when the TX is centrally located. Conversely, a TX positioned off-center exhibits a higher diffusion peak under anistropic diffusion. Furthermore, the propagation of autoinducers (AIs) is inversely proportional to both biofilm size and higher diffusion coefficients along  \(D_\rho\). We demonstrate that the accelerated diffusion of AIs when the TX is off-center is evidenced by anisotropic diffusion from the boundary. This is an evolutionary adaptation that enhances the biofilm's survival. Faster diffusion of AIs and nutrients from the biofilm’s boundary to its core provides protection against hostile environmental changes and ensures the delivery of necessary nourishment.

The findings of this paper present an initial MC model for a 2D biofilm that employs anisotropic diffusion, providing a foundation that closely mimics biofilm under laboratory conditions, thus the groundwork is provided for future research to build on this model. The primary references utilized in this study were accessible; however, there is a need for a standardized nomenclature within the microbiology community for water channels to ensure clarity and consistency. To enhance the QS model in biofilm, further imaging of water channels is necessary. Additionally, incorporating single-particle tracking techniques to observe the movement of molecules in and out of water channels could significantly strengthen and validate the model.

In future work, the accuracy of AI propagation in a real-world biofilm can be enhanced by considering the effects of natural elements (such as change in temperature), as our assumptions are for signal propagation in biofilm under ideal conditions. The effects of the natural elements can severely alter biofilm development and lead to greater heterogeneity of the water channel distribution within real-world biofilms, where diffusion coefficients are likely to vary in both radial and axial directions. Furthermore, the biofilm is modeled as a mature biofilm and overlooks the propagation of AIs throughout the life cycle of the biofilm. Finally, since molecular relaying also takes place in biofilms, future models should consider the inclusion of relaying within the system model. 

\bibliographystyle{IEEEtran}
\bibliography{Reference}

\end{document}